\documentclass[runningheads]{llncs}
\usepackage{graphicx}
\usepackage{xcolor}
\usepackage[colorlinks,linkcolor=blue,citecolor=blue]{hyperref}
\usepackage{multirow,booktabs,amsmath,url,float}
\usepackage[labelfont=bf,labelsep=period]{caption}
\usepackage{subcaption}
\usepackage{pifont} %checkmark Yes and crossmark No
\begin{document}
\title{USC: An Open-Source Uzbek Speech Corpus and Initial Speech Recognition Experiments}
%\title{Towards Multimodal Person Verification Using Audio-Visual-Thermal Streams\thanks{Supported by organization x.}}
%
\titlerunning{USC: An Open-Source Uzbek Speech Corpus}
% If the paper title is too long for the running head, you can set
% an abbreviated paper title here
%
\author{Muhammadjon Musaev\inst{1} \and Saida Mussakhojayeva\inst{2} \and Ilyos Khujayorov\inst{1} \and \\ Yerbolat Khassanov\inst{2} \and Mannon Ochilov\inst{1} \and Huseyin Atakan Varol\inst{2}}
\authorrunning{M. Musaev et al.}
% First names are abbreviated in the running head.
% If there are more than two authors, 'et al.' is used.
%
\institute{Computer Systems, Tashkent University of Information Technology\\ named after Muhammad Al-Khwarizmi, Tashkent, Uzbekistan\\
\email{mm.musaev@rambler.ru, ilyoskhujayorov@gmail.com, ochilov.mannon@mail.ru} \and
Institute of Smart Systems and Artificial Intelligence (ISSAI),\\ Nazarbayev University, Nur-Sultan, Kazakhstan\\
\email{\{saida.mussakhojayeva,yerbolat.khassanov,ahvarol\}@nu.edu.kz}}
\maketitle              % typeset the header of the contribution

\setcounter{footnote}{0} 

\begin{abstract}
We present a freely available speech corpus for the Uzbek language and report preliminary automatic speech recognition (ASR) results using both the deep neural network hidden Markov model (DNN-HMM) and end-to-end (E2E) architectures.
The Uzbek speech corpus (USC) comprises 958 different speakers with a total of 105 hours of transcribed audio recordings.
To the best of our knowledge, this is the first open-source Uzbek speech corpus dedicated to the ASR task.
To ensure high quality, the USC has been manually checked by native speakers.  
We first describe the design and development procedures of the USC, and then explain the conducted ASR experiments in detail.
The experimental results demonstrate promising results for the applicability of the USC for ASR. Specifically, 18.1\% and 17.4\% word error rates were achieved on the validation and test sets, respectively.
To enable experiment reproducibility, we share the USC dataset, pre-trained models, and training recipes in our GitHub repository\footnote{\label{ft:github}\url{https://github.com/IS2AI/Uzbek_ASR}}. 

\end{abstract}
%
%
%
%%%%%%%%%%%%%%%%%%%%%%%%%%%%%%%%%%%%%%%%%%%%%%%%%%%%%%%%%%%%%%%%%%%%%%%%%%%%%%%%%%%%%%%%%%%%%%%%%%%%%%%%%%%%%%%%%%
%%%%%%%%%%%%%%%%%%%%%%%%%%%%%%%%%%%%%%%%%%%%%%%%%%%%%%%%%%%%%%%%%%%%%%%%%%%%%%%%%%%%%%%%%%%%%%%%%%%%%%%%%%%%%%%%%%
\section{Introduction}
In this paper, we present an open-source Uzbek speech corpus (USC) dedicated to advancing automatic speech recognition (ASR) research for the Uzbek language.
Uzbek is the official language of Uzbekistan also spoken in other neighboring countries, such as Afghanistan, Kazakhstan, Kyrgyzstan, Tajikistan, and Turkmenistan.
It is an agglutinative language spoken by over 35 million people worldwide~\cite{UzLangWiki}, which makes it the second-most widely spoken language in the Turkic languages family.
With the USC, we aim to promote the development and usage of the Uzbek language in speech-enabled applications, such as message dictation, voice search, voice command, and other voice-controlled smart devices.
We also believe that the USC will help to facilitate the development of assistive technologies in the Uzbek language for people with special needs (e.g., the hearing impaired).

%One of the main problems with the Uzbek ASR research is the absence of generally accepted common Uzbek corpus.
Previously, several works have addressed Uzbek speech recognition~\cite{musaev2020development,musaev2021automatic}.
%For example, \cite{musaev2021automatic} developed an ASR system for geographical entities using a dataset consisting of 3,500 audio recordings.
%Similarly, \cite{musaev2020development} developed a read speech recognition system trained on the 10 hours of transcribed audio.
However, to the best of our knowledge, there has been no work presenting an open-source Uzbek speech corpus of high quality and sufficient size for training robust speech recognition systems.
%, leading to the absence of generally accepted common Uzbek corpus.
%However, most of them are either publicly unavailable or insufficient for training reliable ASR systems.
As a result, there is no generally accepted common Uzbek dataset, and thus, each research group conducts experiments and reports results on their internal data.
This hinders experiment reproducibility and performance benchmarking, which retards the further development of Uzbek ASR technologies.

To address this problem, we created the USC dataset containing around 105 hours of transcribed audio recordings spoken by 958 speakers from different regions and age groups.
%The dataset comprised of \textbf{X} audio recordings manually transcribed by native speakers.
The USC is primarily designed for the ASR task, however, it can also be used to aid other speech-related tasks, such as speech synthesis and speech translation.
To the best of our knowledge, the USC is the first open-source Uzbek speech corpus available for both academic and commercial use under the Creative Commons Attribution 4.0 International License\footnote{\url{https://creativecommons.org/licenses/by/4.0/}}.
We expect that the USC will be a valuable resource for the general speech research community and become the baseline dataset for Uzbek ASR research.
Therefore, we invite other researchers to use our dataset and help to further explore it with us.

To demonstrate the reliability of the USC, we conducted initial ASR experiments using both the hybrid deep neural network hidden Markov model (DNN-HMM) and end-to-end (E2E) architectures.
Additionally, we investigated the impact of neural language models (LMs) and data augmentation techniques on the Uzbek speech recognition performance.
%We also studied the impact of transfer learning from the Kazakh language, where an ASR system pre-trained on Kazakh speech corpus (KSC)~\cite{khassanov-etal-2021-crowdsourced} is fully fine-tuned using the USC. 
%In our experiments, the best ASR system trained on our dataset achieved X\% and Y\% word error rates (WER) on the validation and test sets respectively, which showcases the high quality of audios and transcripts in USC.
In our experiments, the best DNN-HMM ASR system achieved 18.8\% and 23.5\% word error rates (WER) on the validation and test sets, respectively.
The best E2E ASR system achieved 18.1\% and 17.4\% WERs on the validation and test sets, respectively.
These results showcase the high quality of audios and transcripts in the USC.

The main contribution of this work is two-fold:
\begin{itemize}
    \item We developed the first open-source speech corpus for the Uzbek language.
    \item We conducted initial Uzbek speech recognition experiments using both the conventional DNN-HMM and recently proposed E2E architectures.
%    \item We studied the impact of neural LMs, data augmentation techniques, and transfer learning on the Uzbek speech recognition performance.
\end{itemize}

The rest of the paper is organized as follows:
Section~\ref{sec:related} reviews past works on Uzbek speech recognition and datasets.
Section~\ref{sec:data} extensively describes the USC dataset construction procedures.
The speech recognition experiments and obtained results are presented in Section~\ref{sec:exp}. 
Lastly, Section~\ref{sec:con} concludes the paper and points out directions of future work.

%%%%%%%%%%%%%%%%%%%%%%%%%%%%%%%%%%%%%%%%%%%%%%%%%%%%%%%%%%%%%%%%%%%%%%%%%%%%%%%%%%%%%%%%%%%%%%%%%%%%%%%%%%%%%%%%%%
%%%%%%%%%%%%%%%%%%%%%%%%%%%%%%%%%%%%%%%%%%%%%%%%%%%%%%%%%%%%%%%%%%%%%%%%%%%%%%%%%%%%%%%%%%%%%%%%%%%%%%%%%%%%%%%%%%
\section{Related Works}\label{sec:related}
Speech is the most natural means of communication between humans, and researchers have long dreamed of employing it for interacting with machines.
As a result, ASR research has attracted a great deal of attention over the past few decades~\cite{yu2016automatic}.
In particular, various ASR architectures~\cite{DBLP:journals/spm/X12a,DBLP:conf/icassp/GravesMH13,DBLP:journals/taslp/Abdel-HamidMJDPY14,DBLP:conf/icml/GravesJ14} and annotated datasets~\cite{DBLP:conf/lrec/RousseauDE12,DBLP:conf/icassp/PanayotovCPK15,DBLP:conf/ococosda/BuDNWZ17} for training have been introduced.
Unfortunately, most of the datasets are developed for popular languages such as English, Spanish, and Mandarin whereas less popular languages do not get much attention.
Consequently, the less popular languages face an acute shortage of research and development of ASR technologies~\cite{DBLP:journals/speech/BesacierBKS14a}.
%On the other hand, for the most of the less popular languages the datasets are either too small or does not exist.
%The absence of the reliable datasets is the main factor which prevents the research and development of ASR technology in low-resource languages. 

To address the aforementioned problem, many datasets have been developed in less popular languages.
For example, to advance speech processing research in Kazakhstan, researchers developed open-source Kazakh speech corpora for building speech recognition~\cite{khassanov-etal-2021-crowdsourced} and speech synthesis~\cite{DBLP:journals/corr/abs-2104-08459} applications.
To enable speech research and increase accessibility of speech-enabled applications for illiterate users, Doumbouya et al.~\cite{DBLP:conf/aaai/DoumbouyaEP21} released 150 hours of transcribed audio data for West African languages.
Similarly, several large-scale multilingual speech corpora construction projects were initiated, e.g., VoxForge~\cite{voxforge}, Babel~\cite{DBLP:conf/sltu/GalesKRR14}, M-AILABS~\cite{mailabs}, and Common Voice~\cite{DBLP:conf/lrec/ArdilaBDKMHMSTW20}.
However, these projects do not include the Uzbek language yet.

In the context of the Uzbek language, some works have previously attempted Uzbek speech recognition.
For example, Musaev et al.~\cite{musaev2021automatic} developed an ASR system for geographical entities using a dataset consisting of 3,500 utterances.
Similarly, the authors of~\cite{musaev2020development} developed a read speech recognition system using 10 hours of transcribed audio.
The works of~\cite{DBLP:conf/iscsic/MusaevKO19} and~\cite{musaev2020use} addressed spoken digit and voice command recognition systems under the limited vocabulary scenarios, respectively.
It should be mentioned that the datasets used in these works were very limited and specialized for narrow application domains.
Other existing Uzbek datasets are prohibitively expensive or publicly unavailable~\cite{Speechocean}.
Therefore, the development of an open-source Uzbek speech corpus of sufficient size is of paramount importance. 

%%%%%%%%%%%%%%%%%%%%%%%%%%%%%%%%%%%%%%%%%%%%%%%%%%%%%%%%%%%%%%%%%%%%%%%%%%%%%%%%%%%%%%%%%%%%%%%%%%%%%%%%%%%%%%%%%%
%%%%%%%%%%%%%%%%%%%%%%%%%%%%%%%%%%%%%%%%%%%%%%%%%%%%%%%%%%%%%%%%%%%%%%%%%%%%%%%%%%%%%%%%%%%%%%%%%%%%%%%%%%%%%%%%%%
\section{The USC Dataset Construction}\label{sec:data}
The Uzbek data collection project was conducted with the ethical approval of the Expert Committee consisting of members from the Tashkent University of Information Technology named after Muhammad Al-Khwarizmi.
%and the Ministry for Development of Information Technologies and Communications of the Republic of Uzbekistan.
Each reader participated voluntarily and was informed of the data collection and use protocols.
The dataset was collected by two means: crowdsourcing and audiobooks.
%The dataset construction process consisted of two main stages, i.e. text collection and text narration, which will be thoroughly described in the following sections.

\subsection{Crowdsourcing}
The crowdsourcing process consisted of three main stages--namely, text collection, text narration, and audio checking, which will be thoroughly described in the following sections.

\subsubsection{Text Collection.}
We first collected Uzbek textual data from various sources including news portals, electronic books from modern Uzbek literature and national legislation database.
The texts were collected automatically using web crawlers and they cover a wide range of topics such as politics, finance, entertainment, and law.
In addition, we manually filtered the collected texts to eliminate defects peculiar to web crawlers and exclude non-Uzbek sentences and inappropriate content (e.g., user privacy and violence).
We kept sentences containing borrowed words from other languages such as English.
Lastly, we removed sentences containing numerals and sentences with more than 30 words. 
In total, over 100 thousand sentences were prepared for narration.

\subsubsection{Text Narration.}
\begin{figure}[h]
    %\centering
    \begin{subfigure}[b]{0.45\textwidth}
        \centering
        %\fbox{\includegraphics[width=\textwidth,trim={0.055cm 3.085cm 20.95cm 0.06cm},clip=true]{Figures/telegram_bot1.png}}
        \fbox{\includegraphics[width=\textwidth,trim={0.06cm 0.2cm 19.0cm 0.06275cm},clip=true]{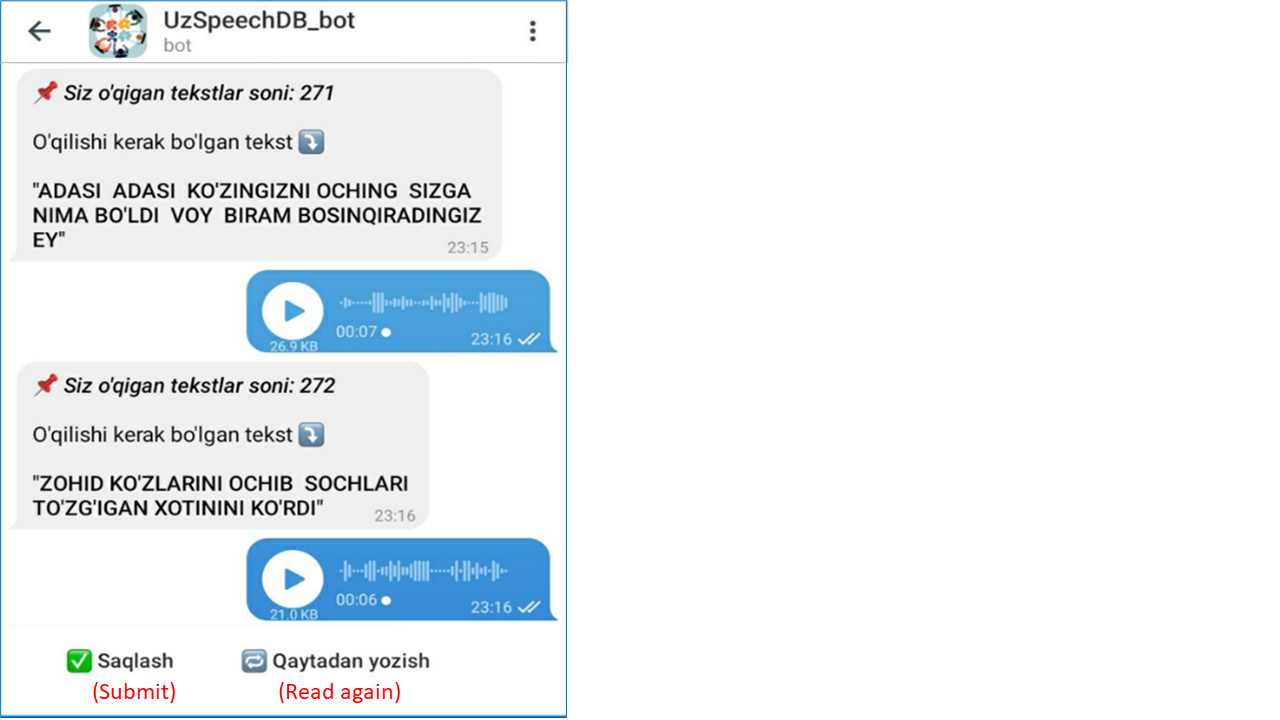}}
        \caption{Data collection bot}
        \label{fig:bot1}
    \end{subfigure}
    %\hfill
    \hspace{0.05\textwidth}
    \begin{subfigure}[b]{0.45\textwidth}
        \centering
        %\fbox{\includegraphics[width=\textwidth,trim={0.09cm 3.15cm 20.975cm 0.091cm},clip=true]{Figures/telegram_bot2.png}}
        \fbox{\includegraphics[width=\textwidth,trim={0.125cm 0.2cm 19.065cm 0.25cm},clip=true]{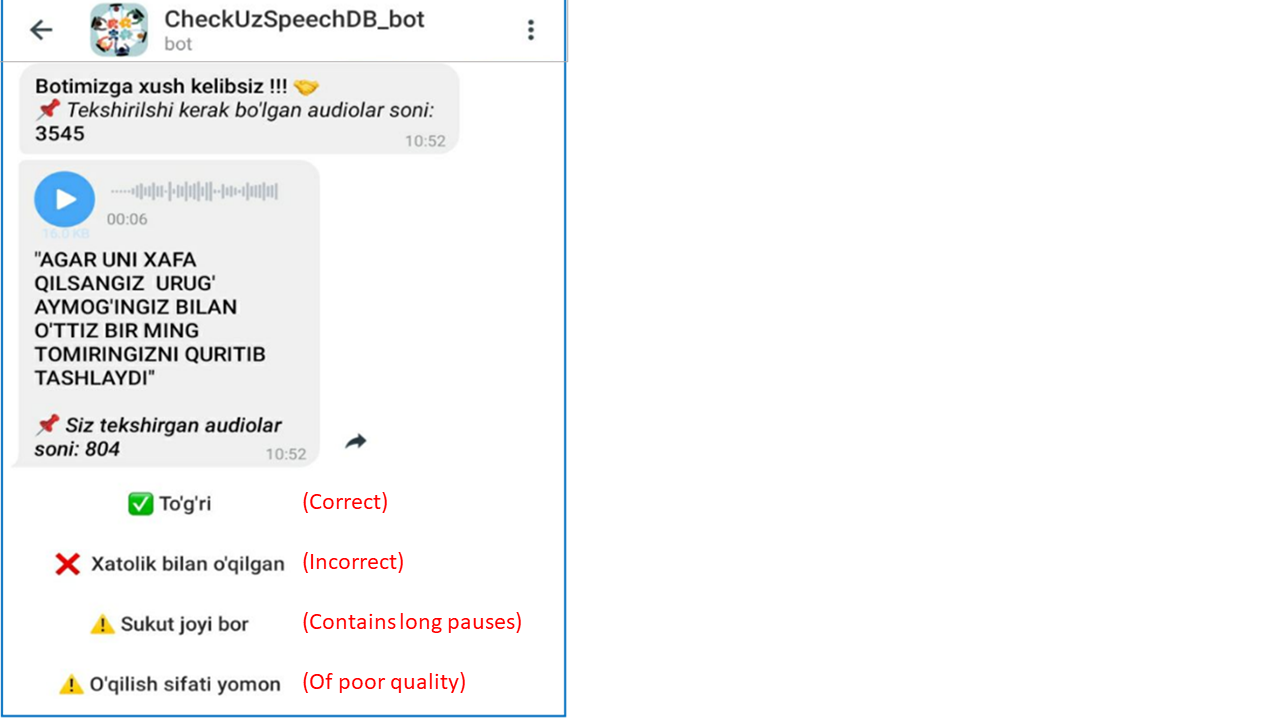}}
        \caption{Data checking bot}
        \label{fig:bot2}
    \end{subfigure}
    \caption{Examples of interaction with the Telegram bots during the (a) data collection and (b) data checking stages.}
    \label{fig:bots}
\end{figure}

To narrate the collected sentences, we employed the Telegram~\cite{telegram} messaging platform, which is widely used in Uzbekistan.
Specifically, we developed a Telegram bot that first presents a welcome message with instructions and then starts the narration process (see Figure~\ref{fig:bot1}).
During the narration process, the bot sends a sentence to a reader and receives the corresponding audio recording.
The bot allowed readers to listen to recorded audio and decide whether to submit or re-record it.
In addition, the bot stored the reader IDs and other information including the age, gender, and geographical location. 
%Thus, incorrectly pronounced sentences can be recorded again.
We attracted readers aged 18 or above by advertising the data collection project in social media, news, and open messaging communities on WhatsApp and Telegram. 

\subsubsection{Audio Checking.}
To ensure the high quality of collected data, we developed an additional Telegram bot for checking the audio recordings.
Different from the audio collection bot, the checker bot sends an audio recording and the corresponding sentence to an examiner (see Figure~\ref{fig:bot2}).
As examiners, we recruited several volunteers among native Uzbek speakers.
The examiners were instructed to inspect received audios with the sentences and mark them as ``correct", ``incorrect", ``contains long pauses", or ``of poor quality".
Audio and sentence pairs marked as ``correct" were added to the final speech corpus.
For pairs marked as ``incorrect", the audio recording was removed, and the sentence was transferred to the audio collection bot for re-reading.
For pairs marked as ``contains long pauses" or ``of poor quality", we manually applied additional quality improvement procedures (e.g., trimming long pauses, splitting audio into several segments, and normalizing audio) and then added the pairs to the final speech corpus.
To make our dataset close to the real-world scenarios, we kept utterances containing background noises.

\subsection{Audiobooks}
To collect data from audiobooks, we extracted freely available audiobooks narrated by 20 Uzbek audiobook narrators.
%In total, we extracted X audiobooks covering different genres from both modern and past authors.
From each book, we took only a 30-minute audio excerpt to balance the data contributed by each speaker.
These excerpts were automatically segmented and aligned with the corresponding text by using the Aeneas Python library~\cite{aeneas}.
The generated segments were manually inspected and then added to the final speech corpus.

\iffalse
To narrate extracted sentences, we prepared voice recording setup and recruited X speaker among native speakers.
The voice recording setup was located in one of our laboratories isolated from the external noises and with the good acoustic properties.
The setup consisted of personal computer and recording device (DEVICE NAME AND MODEL).
The speakers were mostly students, admin and faculty members from our university with the average speaker age X (ranging from Y to Z).
The number of male and female speakers were X and Y, respectively.
The audios were recorded in 48 kHz and 16 bits, but downsampled to 16 kHz and 16 bits for online publication.
\fi

\begin{table}[b]
    \caption{The USC dataset specifications.}\label{tab:stats}
    %\small
    \renewcommand\arraystretch{1.1}
    \setlength{\tabcolsep}{5.75mm}
    \centering
    \begin{tabular}{l|cccc}
        \toprule
        \textbf{Category}       & \textbf{Train}& \textbf{Valid}& \textbf{Test} & \textbf{Total} \\
        \midrule
        Duration (hours)        & 96.4          & 4.0           & 4.5           & 104.9 \\
        \# Utterances           & 100,767       & 3,783         & 3,837         & 108,387 \\
        \# Words                & 569.0k        & 22.5k         & 27.1k         & 618.6k \\
        \# Unique Words         & 59.5k         & 8.4k          & 10.5k         & 63.1k \\
        \# Speakers             & 879           & 41            & 38            & 958  \\
        %\# Speakers             & 788           & 43            & 41            & 872 \\
        \bottomrule
    \end{tabular}
\end{table}

\begin{figure}[th]
  \centering
  \includegraphics[width=1.0\linewidth,trim={0.25cm 9.75cm 9.25cm 0.0cm},clip=true]{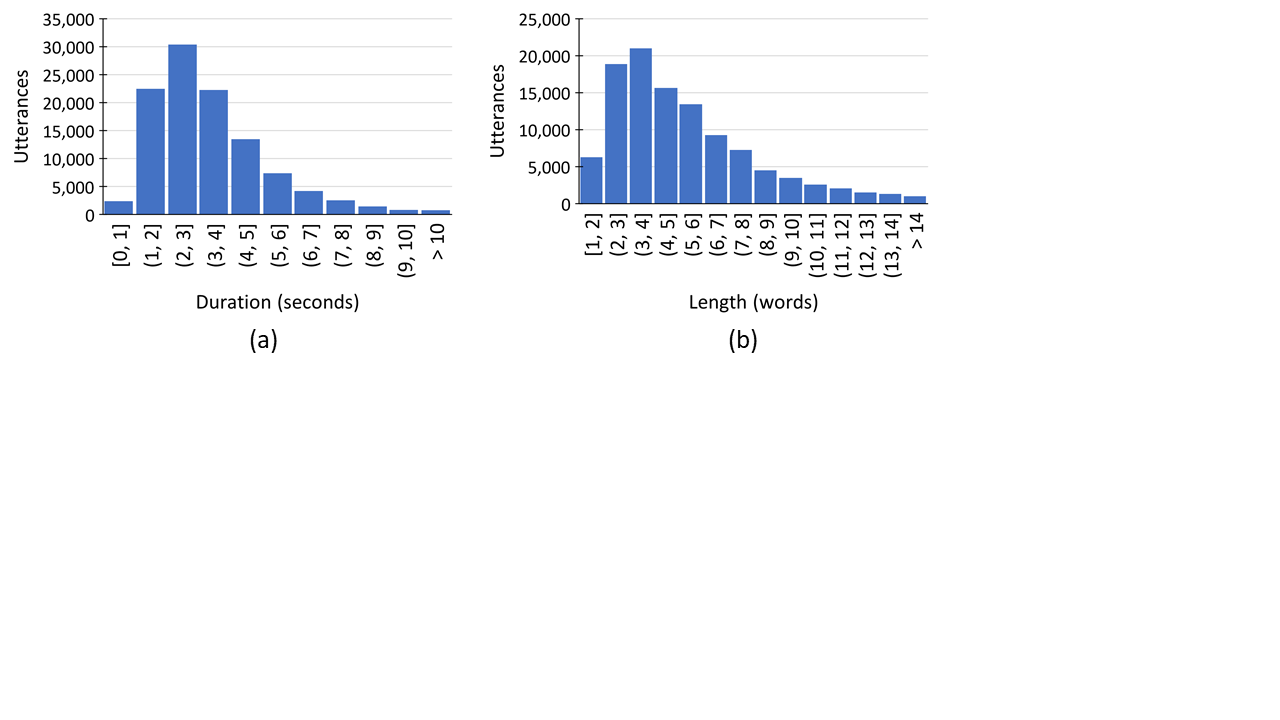}
  \caption{Utterance (a) duration and (b) length distributions in the USC.}
  \label{fig:hist}
\end{figure}

\subsection{Dataset Statistics and Structure}
The dataset statistics are reported in Table~\ref{tab:stats}.
In total, over 108,000 utterances were collected resulting in around 105 hours of transcribed speech data. 
The utterance duration and length distributions are shown in Figure~\ref{fig:hist}.
We split the dataset into training, validation, and test sets.
The speakers in these sets are non-overlapping.
For experiment reproducibility, we ask researchers planning to use our dataset to follow the provided splitting.

The USC dataset is structured as follows.
We split the dataset into three folders corresponding to the training, validation, and test sets.
Each folder contains audio recordings and transcripts.
%The USC dataset consists of audio recordings, transcripts, and metadata stored in separate folders.
The audio and corresponding transcription filenames are the same, except that the audio recordings are stored as WAV files, whereas the transcriptions are stored as TXT files using the UTF-8 encoding.
All the transcriptions are represented using the Uzbek Latin alphabet consisting of 29 letters and the apostrophe symbol.
%The metadata contain the data splitting information for training, validation, and test sets.

%%%%%%%%%%%%%%%%%%%%%%%%%%%%%%%%%%%%%%%%%%%%%%%%%%%%%%%%%%%%%%%%%%%%%%%%%%%%%%%%%%%%%%%%%%%%%%%%%%%%%%%%%%%%%%%%%%
%%%%%%%%%%%%%%%%%%%%%%%%%%%%%%%%%%%%%%%%%%%%%%%%%%%%%%%%%%%%%%%%%%%%%%%%%%%%%%%%%%%%%%%%%%%%%%%%%%%%%%%%%%%%%%%%%%
\section{Speech Recognition Experiments}\label{sec:exp}
We conducted speech recognition experiments to demonstrate the reliability of the USC dataset. We built both DNN-HMM and E2E speech recognition models using our dataset (see Section~\ref{sec:data}) and evaluated them using the character error rate (CER) and word error rate (WER) metrics.
We did not use any external data and other available linguistic resources such as lexicon, pronunciation models, and vocabulary.
%Note that we left the performance comparison of various ASR architectures for Uzbek language as a future work.
Note that we left the detailed performance comparison of various ASR architectures for the Uzbek language as future work.
Hence, in our experiments, we used the standard ASR architectures with the recommended specifications (i.e., number of encoder and decoder blocks, number of layers, layer dimensions, optimizer, initial learning rate, number of training epochs, and so on).
%Therefore, obtained results can't be used to infer the superiority of some ASR architectures over others for Uzbek language.

\subsection{Experimental Setup}
We trained all ASR models using the training set on a single V100 GPU running on the NVIDIA DGX-2 server.
The hyper-parameters were tuned using the validation set, and the best-performing models were evaluated using the test set.
%We didn't apply any word lattice or n-best hypotheses rescoring techniques.
The characteristics of the built DNN-HMM and E2E ASR systems are described in the following sections.
%In particular, we built both the traditional DNN-HMM ASR and recently proposed E2E ASR systems as will be described in the following sections.
%based on the long short-term memory (LSTM)~\cite{DBLP:journals/neco/HochreiterS97}, transformer~\cite{DBLP:conf/nips/VaswaniSPUJGKP17}, and conformer~\cite{DBLP:conf/interspeech/GulatiQCPZYHWZW20} networks as described in the following sections.
%The DNN-HMM ASR models were built using the Kaldi framework~\cite{povey2011kaldi}, whereas the E2E ASR models were built using the ESPnet framework~\cite{watanabe2018espnet}.
For more information on the implementation details and hyper-parameter values, we refer the interested readers to our GitHub repository\textsuperscript{\ref{ft:github}}.
%The interested readers are referred to our GitHube repository\footnote{\url{GitHub_link}}.

\subsubsection{The DNN-HMM ASR.}
%We used Kaldi framework~\cite{povey2011kaldi} to build the DNN-HMM ASR system, and followed the Wall Street Journal (WSJ) recipe.
To build DNN-HMM ASR systems, we used the Kaldi framework~\cite{povey2011kaldi} and followed the Wall Street Journal (WSJ) recipe.
%To build DNN-HMM ASR system, we followed the Wall Street Journal (WSJ) recipe.
%The acoustic model was constructed by stacking 13 time-delay neural network (TDNN)~\citep{DBLP:conf/interspeech/PoveyCWLXYK18} layers with the dimension of 1,024 trained with the lattice-free maximum mutual information (LF-MMI) training criterion~\citep{DBLP:conf/interspeech/PoveyPGGMNWK16}.
The acoustic model was constructed using the factorized time-delay neural networks (TDNN-F)~\cite{DBLP:conf/interspeech/PoveyCWLXYK18} trained with the lattice-free maximum mutual information (LF-MMI)~\cite{DBLP:conf/interspeech/PoveyPGGMNWK16} training criterion.
The inputs were Mel-frequency cepstral coefficients (MFCC) features with cepstral mean and variance normalization extracted every 10 ms over a 25 ms window.
In addition, we applied data augmentation techniques based on the three-way speed perturbation~\cite{DBLP:conf/interspeech/KoPPK15} and spectral augmentation~\cite{DBLP:conf/interspeech/ParkCZCZCL19}.

Due to the strong grapheme-to-phoneme relation in Uzbek, we employed a graphemic lexicon.
The graphemic lexicon is comprised of 59.5k unique words extracted only from the training set.
%The out-of-vocabulary rate in validation and test sets are X\% and Y\%, respectively.
As a language model (LM), we used the Kneser-Ney smoothed 3-gram LM\footnote{We trained several N-gram LMs with different orders and smoothing techniques and picked the one that obtained the best perplexity score on the validation set.} trained on the trascripts of the training set and with the vocabulary covering all words in the graphemic lexicon.

\subsubsection{The E2E ASR.}
To build E2E ASR systems, we used the ESPnet framework~\cite{watanabe2018espnet} and followed the WSJ recipe.
In particular, we built three types of E2E ASR architectures based on the 1) long short-term memory (LSTM)~\cite{DBLP:journals/neco/HochreiterS97}, 2) Transformer~\cite{DBLP:conf/nips/VaswaniSPUJGKP17}, and 3) Conformer~\cite{DBLP:conf/interspeech/GulatiQCPZYHWZW20} networks.
All E2E ASR architectures were jointly trained with the connectionist temporal classification (CTC)~\cite{DBLP:conf/icml/GravesFGS06} objective function under the multi-task learning framework~\cite{DBLP:conf/icassp/KimHW17}.
The input speech features were represented as an 80-dimensional filterbank features with pitch computed every 10 ms over a 25 ms window.
The output units were represented using 29 characters consisting of 26 letters\footnote{Note that the Uzbek Latin alphabet contains 29 letters, however, some of the letters are represented using digraphs (e.g., ng, sh, ch, o' and g'), which we broke down into smaller units and obtained 25 letters. The 26th letter is `w' obtained from international words.}, the apostrophe symbol, and special tokens $<$\textit{unk}$>$ and $<$\textit{space}$>$.
The batch size in all E2E ASR models was set to 64.
To prevent overfitting, we applied data augmentation techniques based on speed perturbation and spectral augmentation.
The results for the Transformer and Conformer based E2E ASR models are reported on the average model constructed using the last 10 checkpoints.

In addition, we built a character-level LSTM LM using the transcripts of the training set.
The LSTM LM was constructed as a stack of two layers each with a memory cell size of 650.
It was employed during the decoding stage using shallow fusion~\cite{DBLP:journals/corr/GulcehreFXCBLBS15} for all the E2E architectures.
%Note that we didn't use any lexicon since our E2E models produce grapheme-based output units.
For decoding, we set the beam size to 20 and the LSTM LM interpolation weight to 1 in all the E2E ASR models.

%Note that we don't need any lexicon for our E2E models since we used the grapheme-based output units.

\textit{1) E2E-LSTM ASR.}
The LSTM-based E2E ASR was constructed using 3 encoder and 1 decoder blocks.
Each encoder block consists of a bidirectional LSTM layer with 1,024 units per direction.
The decoder block consists of a unidirectional LSTM layer with 1,024 units. 
The interpolation weight of the CTC objective was set to 0.5 and 0.3 for the training and decoding stages, respectively.
The model was trained for 100 epochs using the Adadelta optimizer~\cite{DBLP:journals/corr/abs-1212-5701}.

\textit{2) E2E-Transformer ASR.}
The Transformer-based E2E ASR was constructed using 12 encoder and 6 decoder blocks.
We set the number of heads in the self-attention layer to 4 each with 256-dimension hidden states and the feed-forward network dimensions to 2,048.
The interpolation weight for the CTC objective was set to 0.3 for both the training and decoding stages.
The model was trained for 160 epochs using the Noam optimizer~\cite{DBLP:conf/nips/VaswaniSPUJGKP17} with an initial learning rate of 10 and 25k warm-up steps.
The dropout rate and label smoothing were set to 0.1.
%For decoding, we set the beam size to 20 and the LSTM LM interpolation weight to 1.

\textit{3) E2E-Conformer ASR.}
The specifications of the Conformer-based E2E ASR are similar to the Transformer-based model.
It was also constructed using 12 encoder and 6 decoder blocks with a similar number of attention heads and feed-forward network dimensions.
However, the interpolation weight for the CTC objective was set to 0.2 and 0.3 for the training and decoding stages, respectively.
%We set the number of heads in the self-attention layer to 4 each with 256-dimension hidden states and the feed-forward network dimensions to 2,048.
The model was trained for 100 epochs using the Noam optimizer~\cite{DBLP:conf/nips/VaswaniSPUJGKP17} with an initial learning rate of 5 and 25k warm-up steps.
The dropout rate and label smoothing were set to 0.1.

%%%%%%%%%%%%%%%%%%%%%%%%%%%%%%%%%%%%%%%%%%%%%%%%%%%%%%%%%%%%%%%%%%%%%%%%%%%%%%%%%%%%%%%%%%%%%%%%%%%%%%%%%%%%%%%%%%
%%%%%%%%%%%%%%%%%%%%%%%%%%%%%%%%%%%%%%%%%%%%%%%%%%%%%%%%%%%%%%%%%%%%%%%%%%%%%%%%%%%%%%%%%%%%%%%%%%%%%%%%%%%%%%%%%%
\subsection{Experiment Results}
Table~\ref{tab:results} presents the experiment results in terms of the CER and WER on the validation and test sets.
All ASR models achieve competitive results.
Specifically, the best result is achieved by the E2E-Conformer, followed by the E2E-Transformer, the DNN-HMM, and then the E2E-LSTM model.
We observed that integrating LMs into E2E ASR is effective for the Uzbek language, where absolute WER improvements of 7.7\%-12.6\% are achieved on the test set.
The application of speed perturbation to the E2E ASR models gains additional absolute WER improvements of 0.8\%-5.1\% on the test set.
Spectral augmentation further improves the E2E ASR models by absolute WERs of 2.0\%-3.8\% on the test set, however, it does not improve the performance of the DNN-HMM model.
Overall, the lowest WER results are 18.1\% and 17.4\% on the validation and test sets respectively, which were achieved by the E2E-Conformer.
%The data augmentation techniques based on speed perturbation and spectral augmentation are also effective, especially for the E2E models where up to X\% relative WER reduction is achieved.
These results successfully demonstrate the utility of the USC dataset for training ASR models.

\begin{table}[h]
    \caption{The CER (\%) and WER (\%) results of different ASR models built using USC. The impact of language model (LM), speed perturbation (SP), and spectral augmentation (SA) are also reported.}\label{tab:results}
    %\small
    \renewcommand\arraystretch{1.1}
    \setlength{\tabcolsep}{3mm}
    \centering
    \begin{tabular}{l|ccc|cc|cc}
        \toprule
        \multirow{2}{*}{\textbf{Model}} & \multirow{2}{*}{\textbf{LM}}  & \multirow{2}{*}{\textbf{SP}}  & \multirow{2}{*}{\textbf{SA}}  & \multicolumn{2}{|c}{\textbf{Valid}}                                   & \multicolumn{2}{|c}{\textbf{Test}} \\\cline{5-8}
                                        &                               &                               &                               & \textbf{CER}          & \textbf{WER}  & \textbf{CER}                  & \textbf{WER}   \\
        \midrule
        \multirow{2}{*}{DNN-HMM}        & Yes                           & Yes                           & No                            & 6.9                   & 18.8                                         & 7.5                           & 23.5 \\
                                        & Yes                           & Yes                           & Yes                           & 6.9                   & 19.9         & 8.1                           & 24.9 \\\hline
        \multirow{4}{*}{E2E-LSTM}       & No                            & No                            & No                            & 13.8                  & 43.1                                         & 14.0                          & 44.0 \\
                                        & Yes                           & No                            & No                            & 14.9                  & 30.0         & 14.3                          & 31.4 \\
                                        & Yes                           & Yes                           & No                            & 13.7                  & 27.6         & 14.4                          & 30.6 \\
                                        & Yes                           & Yes                           & Yes                           & 12.6                  & 24.9         & 12.0                          & 27.0 \\\hline
        \multirow{4}{*}{E2E-Transformer}& No                            & No                            & No                            & 12.3                  & 35.2                                         & 9.4                           & 31.6 \\
                                        & Yes                           & No                            & No                            & 11.7                  & 25.7         & 8.7                           & 23.9 \\
                                        & Yes                           & Yes                           & No                            & 10.7                  & 23.9         & 8.4                           & 23.0 \\
                                        & Yes                           & Yes                           & Yes                           & 9.9                   & 21.4         & 7.6                           & 21.0 \\\hline
        \multirow{4}{*}{E2E-Conformer}  & No                            & No                            & No                            & 12.7                  & 37.6                                         & 10.7                          & 35.1 \\
                                        & Yes                           & No                            & No                            & 11.5                  & 27.5         & 9.7                           & 26.3 \\
                                        & Yes                           & Yes                           & No                            & 9.2                   & 21.7         & 7.5                           & 21.2 \\
                                        & Yes                           & Yes                           & Yes                           & \textbf{7.8}          & \textbf{18.1}   & \textbf{5.8}          & \textbf{17.4} \\
        \bottomrule
    \end{tabular}
\end{table}

\iffalse
%%%%%%%%%%%%%%%%%%%%%%%%%%%%%%%%%%%%%%%%%%%%%%%%%%%%%%%%%%%%%%%%%%%%%%%%%%%%%%%%%%%%%%%%%%%%%%%%%%%%%%%%%%%%%%%%%%
%%%%%%%%%%%%%%%%%%%%%%%%%%%%%%%%%%%%%%%%%%%%%%%%%%%%%%%%%%%%%%%%%%%%%%%%%%%%%%%%%%%%%%%%%%%%%%%%%%%%%%%%%%%%%%%%%%
\section{Discussion}
\fi

%%%%%%%%%%%%%%%%%%%%%%%%%%%%%%%%%%%%%%%%%%%%%%%%%%%%%%%%%%%%%%%%%%%%%%%%%%%%%%%%%%%%%%%%%%%%%%%%%%%%%%%%%%%%%%%%%%
%%%%%%%%%%%%%%%%%%%%%%%%%%%%%%%%%%%%%%%%%%%%%%%%%%%%%%%%%%%%%%%%%%%%%%%%%%%%%%%%%%%%%%%%%%%%%%%%%%%%%%%%%%%%%%%%%%
\section{Conclusion}\label{sec:con}
We developed an open-source Uzbek speech corpus containing around 105 hours of transcribed audio recordings spoken by 958 speakers.
The corpus was carefully checked by native speakers to ensure high quality.
We believe that our corpus will further advance Uzbek speech processing research and become the primary dataset for comparing different ASR technologies among different research groups.
In addition, we conducted preliminary ASR experiments using both the hybrid DNN-HMM and state-of-the-art E2E architectures.
The best ASR model trained on our dataset achieved 18.1\% and 17.4\% WERs on the validation and test sets respectively, which demonstrates the reliability of the USC.
In future work, we plan to further increase our dataset size and conduct additional ASR experiments.

% ---- Bibliography ----
%
% BibTeX users should specify bibliography style 'splncs04'.
% References will then be sorted and formatted in the correct style.
%
\bibliographystyle{splncs04}
% \bibliography{mybibliography}
%
%\externalbibliography{yes}
\bibliography{main}

\end{document}